\documentclass[usenatbib]{mn2e}
\usepackage{graphicx}
\usepackage{amssymb}
\usepackage{color}

\begin{document}

\newcommand{\mbh}{M_{\bullet}}
\newcommand{\msun}{M_\odot}
\newcommand{\loc}{{\rm l}}



\title[Thermal disc emission: X-ray polarization]
{Thermal disc emission from a rotating black hole: X-ray
polarization signatures}

\author[M.~Dov\v{c}iak, F.~Muleri, R.~W.~Goosmann, V.~Karas and G.~Matt]
{M.~Dov\v{c}iak,\,$^{\!1}$\,\thanks{E-mail: dovciak@astro.cas.cz}
 F.~Muleri,$^{\!2, 3}$
 R.~W.~Goosmann,$^{\!1}$
 V.~Karas$^{1}$ and
 G.~Matt$^{4}$ \\~\\
$^1$~Astronomical Institute, Academy of Sciences of the Czech Republic,
 Bo\v{c}n\'{\i}~II 1401a, CZ-141\,31~Prague, Czech Republic\\
$^2$~Istituto di Astrofisica Spaziale e Fisica Cosmica,
 Via del Fosso del Cavaliere 100, I-00133~Roma, Italy\\
$^3$~Universit\`{a} di Roma Tor Vergata, Dipartimento di Fisica, via
 della Ricerca Scientifica 1, I-00133 Roma, Italy\\
$^4$~Dipartimento di Fisica, Universit\`a degli Studi ``Roma Tre'',
 Via della Vasca Navale 84, I-00146~Roma, Italy}

\date{Accepted .... Received ...}
\pagerange{\pageref{firstpage}--\pageref{lastpage}}
\pubyear{2007}
\maketitle
\label{firstpage}

\begin{abstract}
Thermal emission from the accretion disc around a black hole
can be polarized, due to Thomson scattering in a disc atmosphere.
In Newtonian space, the polarization angle must be either parallel
or perpendicular to the projection of the disc axis on the
sky. As first pointed out by Stark and Connors in 1977, General Relativity
effects strongly modify the polarization properties of the thermal
radiation as observed at infinity.
Among these effects, the rotation of the polarization angle
with energy is particularly useful as a diagnostic tool.

In this paper, we extend the Stark and Connors calculations by including
the spectral hardening factor, several values of the optical depth of the
scattering atmosphere and rendering
the results to the expected performances of planned X-ray polarimeters.
In particular, to assess the perspectives for the next generation
of X-ray polarimeters, we consider the expected sensitivity of the detectors
onboard the planned POLARIX and IXO missions.
We assume the two cases of a Schwarzschild and an extreme Kerr
black hole with a standard thin disc and a scattering atmosphere. We
compute the expected polarization degree and the angle as functions
of the energy as they could be measured for different inclinations of
the observer, optical thickness of the atmosphere and different values
of the black hole spin.
We assume the thermal emission dominates the X-ray band.
Using the
flux level of the microquasar GRS 1915+105 in the thermal state, we calculate
the observed polarization.
\end{abstract}


\section{Introduction}
The imprints of General Relativity (GR) effects on the radiation
emitted in the inner regions of the accretion disc around the black
hole in Active Galactic Nuclei (AGN) and Galactic black hole systems
(GBHS) have been the subject of intense observational efforts in the
recent past. The iron line spectroscopy has been so far the most used and
successful technique to study GR effects and to estimate the black
hole spin in both classes of objects (see \citealt{Miller07} for a
recent review and references therein). Recently, continuum
measurements have also been shown to be a promising technique for GBHS
(e.g. \citealt{Liu}).

X-ray polarimetry is also a potentially very
powerful technique in this respect, as shown by \citet{mat93} and
Dov\v{c}iak, Karas \& Matt (\citeyear{dovc04b}) for AGN. For GBHS,
\citet{sc77}; \citet{cs77} and Connors, Stark \& Piran
(\citeyear{con80}) discussed the GR
effects on the polarization properties of the thermal emission of the
accretion disc.  They showed that a strong variation of the
polarization angle and degree with energy, due to the radial
dependence of the temperature, is induced.
In this paper we reconsider and extend the Stark and Connors calculations,
mainly we consider the scattering atmosphere with different optical depths.

We assume the Kerr metric for the gravitational field, a standard thin
disc with a scattering atmosphere
for the accretion flow and Thomson scattering in the disc atmosphere as the
origin of polarization.
The resulting polarization angle and degree will be shown as functions of the
energy for different inclinations of the observer, optical thickness of the
atmosphere and black hole spins.

The advent of a new generation of X-ray polarimeters
(\citealt{Costa2001}) makes the observation of the effects discussed
in this paper feasible.  We will discuss the observational
perspectives for a bright GBHS, the well-known high inclination
microquasar GRS~1915+105, for both a relatively small polarimetric
mission like POLARIX (\citealt{Costa2006}) and for the polarimeter
aboard IXO (\citealt{Bellazzini2006b}).

\section{Assumptions of the model}
We assume a Keplerian, geometrically thin and optically thick disc
around a black hole (Shakura \& Syunyaev \citeyear{shak73}).
A full General Relativity treatment is adopted (Page \& Thorne
\citeyear{page74}).
We will show results for both a Schwarzschild ($a=0$) and an
extreme Kerr ($a=1$) black hole, where $a$ is the
dimensionless angular momentum per unit mass of the black hole. At each
radius the disc is assumed to emit as a black body, modified by
Thomson scattering (see below).
The radial dependence of the temperature is given by the
Novikov-Thorne relativistic expression to which we
add the hardening factor $f_{\rm col}$ correction,
\begin{equation}
\label{temp}
T(r,a) \approx 741\,f_{\rm col}\,\left[\frac{\dot{M}}{\msun/y}\right]^{1/4}\,
         \left[\frac{\mbh}{\msun}\right]^{-1/2}\,R(r,a)\quad
\end{equation}
where $R(r,a)=r^{-3/4}\,[\mathcal{L}(r,a)/\mathcal{C}(r,a)]^{1/4}$
and $\mathcal{L}(r,a)$ and $\mathcal{C}(r,a)$ are functions of the radius
and black hole spin (see Novikov \& Thorne \citeyear{novi73} for the details).
We suppose that there is zero torque at the
inner edge of the disc, which is assumed to be coincident with the
innermost stable circular orbit, i.e., $r_{\rm ISCO}=6\,r_{\rm g}$ (for
Schwarzschild black hole) and $r_{\rm ISCO}=1\,r_{\rm g}$ (for an extreme
Kerr black hole) with $r_{\rm g}=GM/c^2$.

\begin{figure}
\begin{center}
  \vspace*{3.45mm}
  \includegraphics[height=5cm]{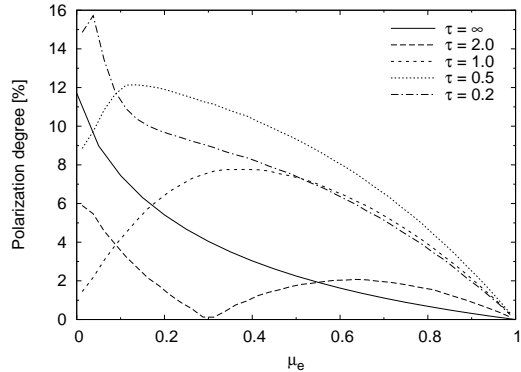}
  \caption{The dependence of the local polarization degree on the cosine of the
emission angle. Curves for different optical depths $\tau$ are shown. The results
for $\tau  \gtrsim 5$ do not differ much from the case $\tau=\infty$.}
  \label{fig-pda_loc}
\end{center}
\end{figure}

We also assume that photons are scattered in the atmosphere of the
disc and thus the observed radiation is polarized. We assume
Thomson scattering and no absorption in the atmosphere.
The resulting polarization
in the local reference frame of the disc atmosphere is computed using
the Monte Carlo code {\sc STOKES} \citep{goos07} for finite
optical depths; for infinite optical depths we adopt Chandrasekhar's
analytical formula (Chandrasekhar \citeyear{chan60}).
In Fig.~\ref{fig-pda_loc}, the polarization
degree we obtain for the local disc emission is plotted versus
the cosine of the emission angle, $\mu_e$, and for different vertical
optical Thomson depths, $\tau$, of the scattering atmosphere. For
$\tau \lesssim 1$ the polarization vector of the emission is
aligned with the projected symmetry axis of the accretion disc. At
higher optical depths, the direction of the polarization vector
switches to a perpendicular one with respect to the symmetry
axis. The intermediate case with $\tau = 2$ has parallel polarization for
$\mu_e > 0.3$ and perpendicular polarization otherwise.

\begin{figure}
\begin{center}
\begin{tabular}{c}
  $\phantom{\hspace*{9mm}}\tau=\infty$\\
  \includegraphics[height=5cm]{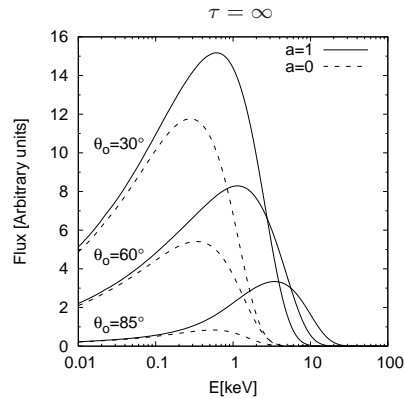}
\end{tabular}
  \caption{The multicolour black body spectra (energy flux $f_{\rm o}E$) for the
  extreme Kerr (solid line)   and Schwarzschild (dashed line) black hole for the
  three observer inclinations of
  30$^\circ$, 60$^\circ$ and 85$^\circ$ and for the semi-infinite atmosphere.}
  \label{fig-flux}
\end{center}
\end{figure}

The effect of hardening of the energy of photons
due to scattering is modelled by a hardening factor, which increases
the effective temperature (Shimura \& Takahara \citeyear{shim95}). In
Fig.~\ref{fig-flux} we show the multicolour black-body energy flux integrated over
the
whole disc as it would be measured by a distant observer. The normalization of
the flux scales with the cosine of the inclination angle.  For the extreme Kerr
black hole, the disc can  extend closer to the black hole horizon
and therefore reach higher temperatures. Therefore, the emission is
stronger and harder in this case. Here, and in all computations we assume
the mass of the central black hole $\mbh=14\msun$, the accretion rate $\dot
M=1.4\times10^{18}\,$g/s and the hardening factor $f=1.7$. With this choice,
the maximum temperature in the disc  is 0.389 keV at the radius $10\,r_{\rm g}$ in
the Schwarzschild case and 1.985 keV at $1\,r_{\rm g}$ in the extreme Kerr case.

The assumption of a pure scattering atmosphere is the most critical
one as far as the polarization is concerned.
If the absorption opacity of the atmosphere is not negligible,
the polarization degree is reduced and the polarization angle, for large
optical depths, can even switch from perpendicular (as for a pure
scattering atmosphere) to parallel to the projection of the axis on the sky
(e.g. Laor, Netzer \& Piran \citeyear{lao90}; Matt, Fabian \& Ross
\citeyear{mfr93}). Our assumption
is therefore optimistic; a more realistic calculation would need a
detailed modelling of the disc atmosphere. We stress here that, whatever the
absorption opacity is, in the Newtonian case (i.e. when the effects of special
and general relativity are neglected) the polarization,
for symmetry reasons, must be either parallel or perpendicular to the
projection of the disc axis onto the sky. The continuous variation
of the polarization angle with energy (see the next section) is a clear
signature of the GR effects.

Once the photons leave the atmosphere the polarization vector can be
rotated due to the strong gravity of the central black hole. The
emission is amplified by the transfer function (Cunningham
\citeyear{cunn75}; Dov\v{c}iak, Karas \& Yaqoob \citeyear{dovc04c}) and the
energy of photons is shifted by the gravitational and Doppler
effects.

We assume that scattering occurs in a
non-expanding atmosphere, so we use the transfer function computed for
the equatorial plane and we do not include the additional rotation of
the polarization angle that could be caused by the bulk outflow
velocity (see Beloborodov \citeyear{belo98}).

Also one can expect that magnetic fields of the disc atmosphere (Silantev
\& Gnedin \citeyear{sila08}) will interfere the observed polarization. However,
the energy-resolved dependency of the thermal component exhibits a rather
unique profile which distinguishes the gravitational effects from other
influences.

As the most promising polarimeters presently designed work in X-rays
above a few keV, in this paper we limit ourselves to investigate the
case of a stellar mass black hole (for AGN, the thermal emission is
dominant in the UV band).

\section{Numerical implementation}
The main current application of our
code is within the {\sc xspec} data-fitting package, however, its
applicability goes beyond that: the code can be compiled in a
stand-alone mode, capable of examining polarimetrical quantities
in the strong-gravity regime (see Dov\v{c}iak \citeyear{dovc04a};
Dov\v{c}iak et al.\ \citeyear{dovc04c,dovc04d}).

Properties of radiation are described in terms of photon numbers.
For a point-like source, the measured flux in the solid angle ${\rm d}
\Omega_{\rm o}$ is associated with the detector area
${\rm d}S_{\rm o}{\equiv}D^2\,{\rm d}\Omega_{\rm o}$. This relation
defines the distance $D$ between the observer and the source.
The observed photon flux per unit solid angle in the energy bin
$\langle E, E+\Delta E\rangle$,
$\Delta f_{\rm o}\equiv{{\rm d}N}/{{\rm d}t\,{\rm d}\Omega_{\rm o}}$,
is
\begin{equation}
\label{emission1}
\Delta f_{\rm o}
=\int_{r_{\rm in}}^{r_{\rm out}}\!\!
{\rm d}r\,r\,\int_{\phi}^{\phi+{\Delta\phi}}\!\!{\rm d}\varphi\,
\int_{E/g}^{(E+\Delta E)/g}\!\!{\rm d}E_\loc\,f_{\loc}\,G,
\end{equation}
where $G\equiv g^2 l\mu_{\rm e}$ is the transfer function with
$g,\,l$ and $\mu_{\rm e}$ being the energy shift, lensing and cosine of the emission
angle, respectively, and
\begin{equation}
f_\loc\equiv
\frac{{\rm d}N_{\loc}}{{\rm d}\tau\,{\rm d}S_{\loc}\,
{\rm d}\Omega_{\loc}\,{\rm d}E_{\loc}}
\end{equation}
is a local photon flux emitted from the surface of the disc.
For a derivation of the equation (\ref{emission1}) and a more detailed description
of numerical computations of the transfer function see Dov\v{c}iak
(\citeyear{dovc04a}) and Dov\v{c}iak et al.\ (\citeyear{dovc04d}).

Let us define the specific Stokes parameters,
\begin{equation}
i_\nu\equiv \frac{I_{\nu}}{E}\, ,\quad q_\nu\equiv \frac{Q_{\nu}}{E}\, ,\quad
u_\nu\equiv \frac{U_{\nu}}{E}\, ,\quad v_\nu\equiv \frac{V_{\nu}}{E}\, ,
\end{equation}
where $I_{\nu}$, $Q_{\nu}$, $U_{\nu}$ and $V_{\nu}$ are Stokes
parameters for light with frequency $\nu$
(further on, we drop the index $\nu$ but we will always
consider these quantities for light of a given frequency).

The integrated specific Stokes parameters (per energy bin
$\langle E, E+\Delta E\rangle$), i.e.\ $\Delta
i_{\rm o}$, $\Delta q_{\rm o}$, $\Delta u_{\rm o}$ and $\Delta v_{\rm
o}$ are the quantities that the observer determines from the local
specific Stokes parameters $i_\loc$, $q_\loc$, $u_\loc$ and $v_\loc$ on the
disc in the following way:
\begin{eqnarray}
\label{S1}
{\Delta}i_{\rm o} & = & \int{\rm d}S\,\int{\rm d}E_{\loc}\,
i_{\loc}\,G\, ,\\
\label{S2}
{\Delta}q_{\rm o} & = & \int{\rm d}S\,\int{\rm d}E_{\loc}\,
[q_{\loc}\cos{2\Psi}-u_{\loc}\sin{2\Psi}]\,G\, ,\\
\label{S3}
{\Delta}u_{\rm o} & = & \int{\rm d}S\,\int{\rm d}E_{\loc}\,
[q_{\loc}\sin{2\Psi}+u_{\loc}\cos{2\Psi}]\,G\, ,\\
\label{S4}
{\Delta}v_{\rm o} & = & \int{\rm d}S\,\int{\rm d}E_{\loc}\,
v_{\loc}\,G\, .
\end{eqnarray}
Here, ${\rm d}S=r\,{\rm d}r\,{\rm d}\varphi$ is the integration element of the
coordinate area on the disc and $\Psi$ is a change of the polarization angle.
The integration boundaries are the
same as in eq.~(\ref{emission1}). As can be seen from the
definition, the first specific Stokes parameter is equal to the photon
flux, therefore, eqs.~(\ref{emission1}) and (\ref{S1}) are
identical.

In eqs.\ (\ref{S1})--(\ref{S4}) we used a law of transformation of the Stokes
parameters by the rotation of axes, see eqs.~(I.185) and (I.186) in
Chandrasekhar (\citeyear{chan60}).

An alternative customary way of expressing polarization is in terms of
degree of polarization $P_{\rm o}$ and two polarization angles
$\chi_{\rm o}$ and $\xi_{\rm o}$, defined by
$P_{\rm o}  =  \sqrt{q_{\rm o}^2+u_{\rm o}^2+v_{\rm o}^2}/i_{\rm o}$,
$\tan{2\chi_{\rm o}} = u_{\rm o}/q_{\rm o}$, and
$\sin{2\xi_{\rm o}} = v_{\rm o}/\sqrt{q_{\rm o}^2+u_{\rm o}^2+v_{\rm o}^2}.$
However, these relations can be used only when we are interested in energy
dependence of polarization. If we want to compute the polarization in certain
broad energy range we need to integrate the Stokes parameters and use them instead
of the specific Stokes parameters to get the correct result.

Various physical effects can influence polarization of light as it propagates
towards an observer. Here we examine only the influence of the gravitational
field represented by the vacuum Kerr space-time.
The change of the polarization angle $\Psi$ is defined as the angle by which
a vector parallelly transported along the light geodesic rotates with respect
to some chosen frame. We define it in this
way because in vacuum the polarization vector is parallelly transported along
the light geodesic. This angle depends on the choice of the local frame at
the disc and at infinity. At the disc we consider the local frame co-moving with
it with the $x$-axis in the direction $-\partial/\partial\theta$ in the
plane defined
by the normal of the disc $n^\mu$ and the momentum $p_{\rm e}^\mu$ of the emitted
photon. It is perpendicular to the momentum $p_{\rm e}^\mu$. The $y$ axis lies in
the plane of the disc, is perpendicular to the momentum $p_{\rm e}^\mu$ and has
direction $-\partial/\partial\varphi$. At infinity we consider static frame attached
to the observer's sky with $x$-axis identified with the impact parameter $\beta$
(defined as positive in the direction $-\partial/\partial\theta$) and $y$-axis
identified with the impact parameter $-\alpha$ ($\alpha$ is defined as positive in
the direction $\partial/\partial\varphi$).

\begin{figure*}
\includegraphics{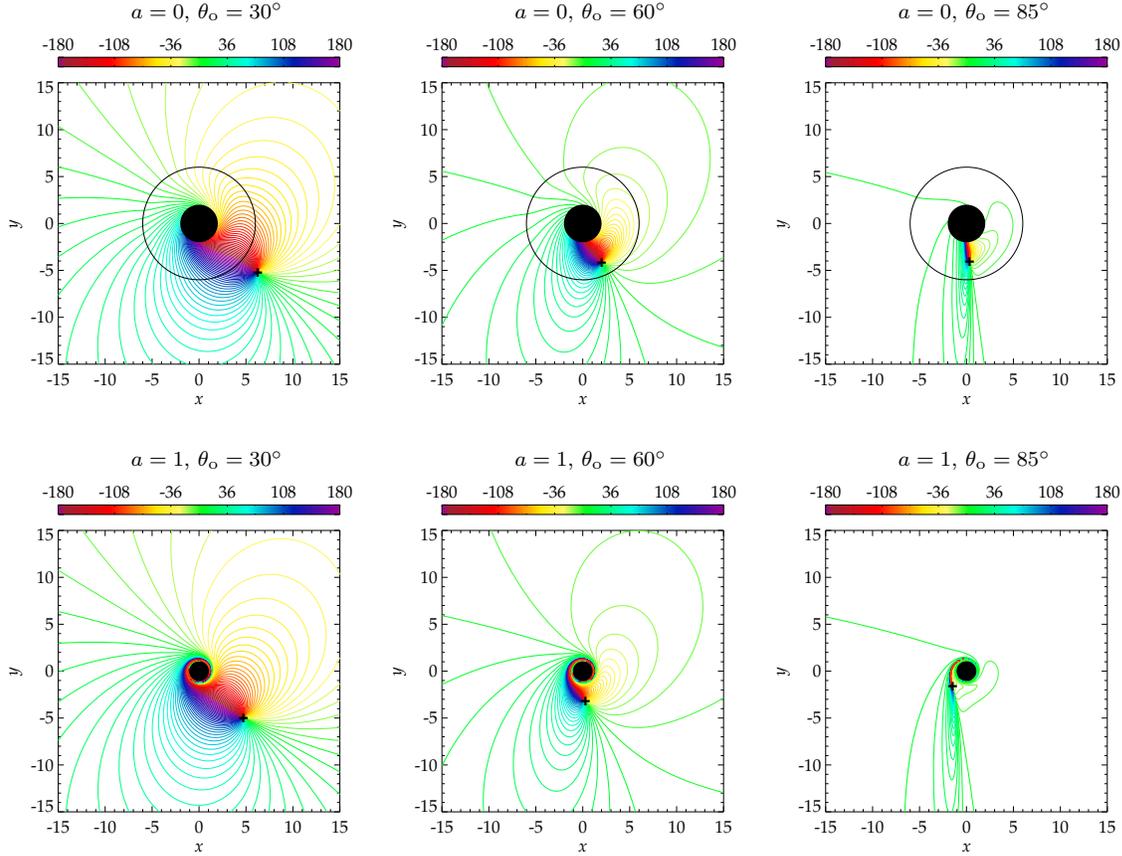}
\caption{Contour graphs of the change of the polarization angle for
Schwarzschild (top) and extreme Kerr (bottom) black hole. The observer
inclination is $\theta_{\rm o}=30^\circ,\, 60^\circ$ and $85^\circ$ (left to
right). The observer is located to the top of the pictures. The innermost
stable circular orbit is shown for the Schwarzschild case. The critical point,
where the photons are emitted perpendicularly to the disc, is shown by a cross.
The black hole rotates counter-clockwise in the Kerr case. The graphs are
represented in the coordinates $x=r\cos{\varphi},\,y=r\sin{\varphi}$ in the
equatorial plane where $r$ and $\varphi$ are Boyer-Lindquist coordinates.}
\label{fig-cpa}
\end{figure*}

The change of polarization angle $\Psi$ is (see Connors et al.\ \citeyear{con80})
\begin{equation}
\label{polar}
\tan{\Psi}=\frac{Y}{X}\, ,
\end{equation}
where
\begin{eqnarray}
X & = & -(\alpha-a\sin{\theta_{\rm o}})\kappa_1-\beta\kappa_2\, ,\\
Y & = & \phantom{-}(\alpha-a\sin{\theta_{\rm o}})\kappa_2-\beta\kappa_1\, ,
\end{eqnarray}
with the black hole spin $a$ positive when the black hole rotates counter-clockwise,
i.e. in the direction $\partial/\partial\varphi$. The angle $\theta_{\rm o}$ is
the observer's inclination, $\kappa_1$ and $\kappa_2$ are components of the
complex Penrose-Walker constant of parallel transport along geodesic
$\kappa_{\rm pw}$ (Walker \& Penrose \citeyear{wal70})
\begin{eqnarray}
\nonumber
\kappa_1 \hspace*{-2mm} & = & \hspace*{-2mm} arp_{\rm e}^\theta f^t-r\,
[a\,p_{\rm e}^t-(r^2+a^2)\,p_{\rm e}^\varphi]f^\theta\\[1mm]
& & \hspace*{-2mm} -r(r^2+a^2)\,p_{\rm e}^\theta f^\varphi\, ,\\[2mm]
\kappa_2 \hspace*{-2mm} & = & \hspace*{-2mm} -r\,p_{\rm e}^rf^t+r\,
[p_{\rm e}^t-a\,p_{\rm e}^\varphi]f^r+arp_{\rm e}^rf^\varphi\, .
\end{eqnarray}
Here, the polarization vector $f^\mu$ is
\begin{equation}
f^\mu = \frac{n^\mu-\mu_{\rm e}\left( g\,p_{\rm e}^\mu-U^\mu\right)}
{\sqrt{1-\mu_{\rm e}^2}}\, .
\end{equation}
with $U^\mu$ being the four-velocity of the co-rotating accretion disc\footnote{
Below the innermost stable circular orbit we consider a free fall with the
conserved energy and angular momentum of the matter on the marginally stable
orbit. Note however, that we use this region only when computing the change of
the polarization angle on Fig.~\ref{fig-cpa}. The thermal emission is neglected
in this region.}.
Thus it is chosen to be in the direction of the $x$-axis in the above defined
local frame on the disc.

\section{Polarization at detector} \label{sec:Polarization}

The local polarization induced by scattering (see
Fig.~\ref{fig-pda_loc}) is defined in the local frame co-moving with
the disc.  The intrinsic polarization angle is perpendicular to the
disc axis for  large optical depths and parallel for small
optical depths. Photons are travelling in a curved space-time, which
changes their polarization angle as seen at infinity, eq.~(\ref{polar}),
depending on the particular trajectory of each photon.  This results in a
depolarizing effect, when all photons will be eventually summed up at
infinity, see eqs.~(\ref{S1})--(\ref{S4}).

\begin{figure*}
\hspace*{-1mm}\includegraphics{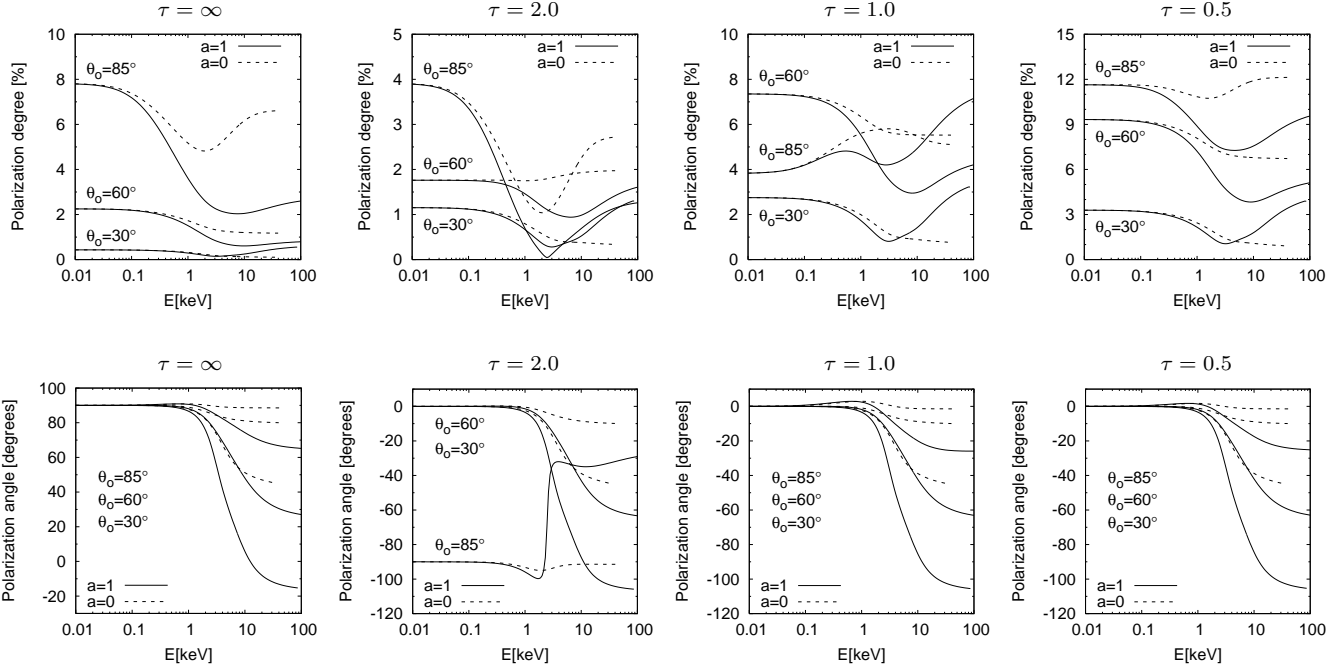}
\caption{The energy dependence of the polarization degree (top) and
  polarization angle (bottom) as the observer at infinity would measure
  them for different optical depths of the atmosphere
  $\tau=\infty,\,2.0,\,1.0,\,0.5$ (from left to right), in the case of
  the Schwarzschild (dashed lines) and extreme Kerr (solid lines)
  black holes and observer inclinations of 30$^\circ$, 60$^\circ$ and
  85$^\circ$. A polarization angle of $0\degr$ represents the
  direction aligned with the projected symmetry axis of the disc.}
\label{fig-pda1}
\end{figure*}

\begin{figure*}
\hspace*{-1mm}\includegraphics{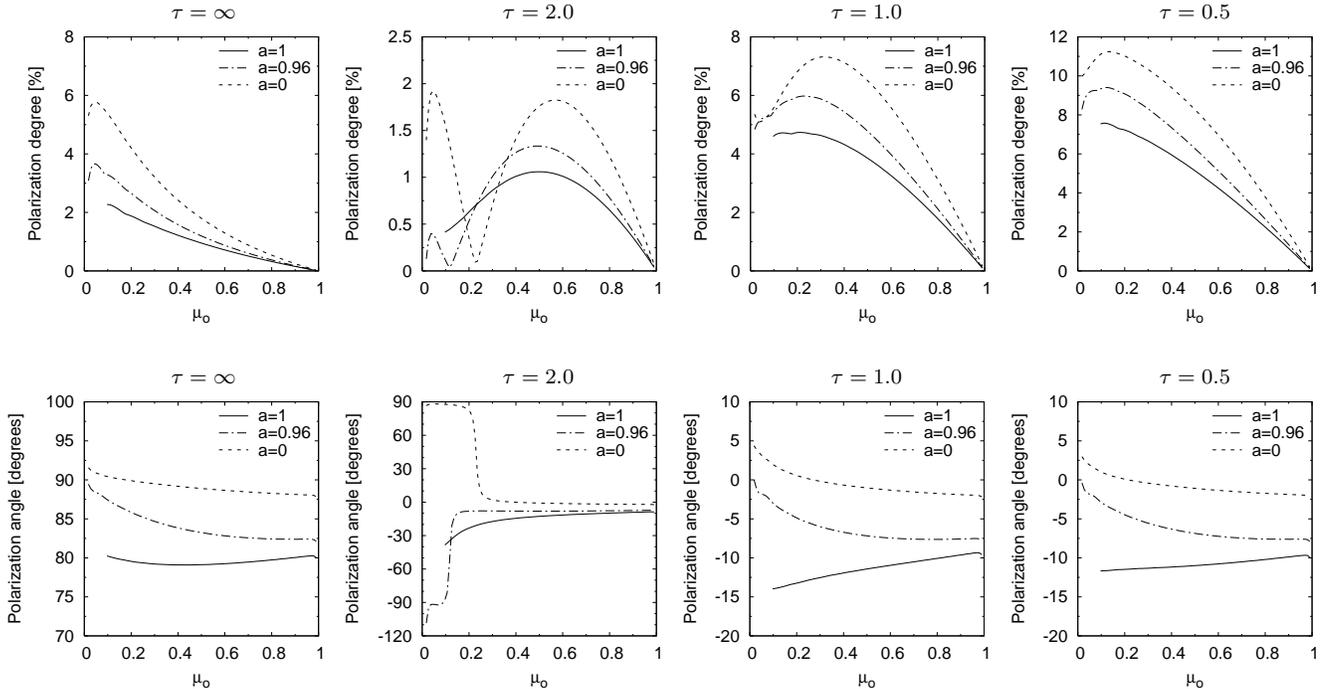}
\caption{The dependence of the polarization degree (top) and
polarization angle (bottom) at infinity on cosine of the observer's
inclination for different optical depths of the atmosphere
$\tau=\infty,\,2.0,\,1.0,\,0.5$ (from left to right), in the case of
the Schwarzschild (dashed lines) and extreme Kerr (solid lines) black
holes. A polarization angle of $0\degr$ represents the direction
aligned with the projected symmetry axis of the disc.}
\label{fig-pda2}
\end{figure*}

Due to aberration the change of the polarization angle for each
photon is different at infinity even in the Schwarzschild space-time
(\citealt{pin77}). In the Kerr case the rotation of polarization vector
because of the gravitational dragging is added.  The dependence of the
change of the polarization angle on the position of the emission from
the disc is shown in Fig.~\ref{fig-cpa}. It can be seen that at large
distances from the black hole this change is small. Therefore the
depolarizing effect of the integration over this part of the disc will
be also relatively small. On the other hand, below the critical point,
where the light is emitted perpendicularly to the disc
(see Fig.~\ref{fig-cpa}),
the change of the polarization angle can acquire any value.
Thus the depolarizing effect of this region is
large. Therefore for the overall polarization measured at infinity it
is important if this critical point is above or below the marginally
stable orbit.

It is clear that the depolarizing region is smaller
for lower spin of the black hole. Note also that the area of the
depolarizing region is larger for lower inclinations of the observer
(the critical point moves farther away from the black hole).


The decrease of the temperature with the disc radius implies that, the
closer to the black hole the radiation is emitted, the harder it
is. As a consequence, the effects of  GR are larger at higher
energies.  The energy dependence of the polarization degree and angle
at infinity is shown in Fig.~\ref{fig-pda1}. In the low energy
limit, i.e. for $E\lesssim 0.1$~keV, where the radiation is
emitted far away from the black hole, the polarization degree and
angle are basically equal to the local ones for the emission angle
equal to the inclination of the observer (compare with the
Fig.~\ref{fig-pda_loc}).  In general, the polarization is highest for
these energies because, as said above, GR effects tend to depolarize
the radiation emitted from regions closer to the central black hole.

For high energies (above 10 keV), the polarization degree also
increases.  The polarization here is influenced mainly by that region
of the disc, where the transfer function is the largest and the
temperature is the highest.  This area is not very large and thus the
span of the change of the polarization angle is small. Therefore the
depolarizing effect is not very large. Note, however, that for the
same reason (small area) the flux for this interval of energy is also
very small (see Fig.~\ref{fig-flux}).

The dependence of the polarization degree and angle on the inclination
of the observer is shown in Fig.~\ref{fig-pda2}. Both quantities are
integrated over the whole energy range (0.01--100 keV). The
polarization degree in most cases increases with the inclination angle
of the observer. The polarization angle does not change much with the
observer's inclination and optical depth, apart from the highest
inclinations.  Its value depends, quite strongly, on the spin of the black
hole.

\section{Observational perspective}

The development of X-ray polarimeters based on the photoelectric
effect has renewed the interest in the polarimetry of X-ray
astronomical sources.  Until now, the only positive detection is that
of the Crab Nebula (Weisskopf et al. \citeyear{Weisskopf1978}). The
\emph{Gas Pixel Detector} (GPD, Costa
et al. \citeyear{Costa2001}; Bellazzini et al. \citeyear{Bellazzini2006a},
\citeyear{Bellazzini2007}) is one of
the most advanced project in this field. Its current design allows a
large increase of  performances with respect to the instruments built
so far, with a sensitivity peaking at about 3 keV and a response
between 2 and 10~keV.

The GPD has been proposed as a focal plane instrument, coupled with an
X-ray optics,  for several satellite missions. Here we discuss two
feasible scenarios:

\begin{itemize}
\item a relatively small, low-cost pathfinder mission, possibly
dedicated to the X-ray polarimetry, which could perform the study of
Galactic and bright extragalactic sources. Long pointed observations
are possible in this scenario and polarimetry at the level of
$\sim$1\% could be reached in a few days of observation, depending on
the flux of the source;
\item a large observatory, which could be dedicated to the study of
the faint extragalactic sources and to the detailed investigation of
the most interesting ones, following the results from the pathfinder
missions. In this scenario, the lower time dedicated to the X-ray
polarimetry will be balanced by the larger area of the optics.
\end{itemize}

We use two examples from the two scenarios above to verify if the
effects presented in Sec.~\ref{sec:Polarization} will be measurable by
the next missions with the GPD on-board. In the former profile,
various missions are currently under study and hence there are
concrete possibilities that at least one mission with an X-ray
polarimeter based on the GPD will be launched in a few years. One of
the options is POLARIX (Costa et al.  \citeyear{Costa2006}), an Italian
mission dedicated to the X-ray polarimetry.  Its design is now under Phase~A
study by the Italian Space Agency and, if eventually
selected, POLARIX will be launched between 2012 and 2014. The baseline
assumes three optics and even an improved version with six telescopes
is considered. As a benchmark for the pathfinder missions, we considered this
improved design of POLARIX, with the warning
that, for the baseline version, the same results can be achieved by
doubling the observing time.

With regard to the large mission scenario, the GPD was also inserted in
the focal plane of the XEUS mission (Bellazzini
et al.  \citeyear{Bellazzini2006b}), which was proposed to
ESA for the Cosmic Vision 2015-2025 and has passed the first selection
step. It is now considered as a possible focal plane instrument for the
International X-ray Observatory (IXO) mission, arising from the merging
of the former XEUS and Constellation-X missions. Because the details of the
IXO mission have not been fixed yet, we use the XEUS design (e.g.
Arnaud et al. \citeyear{Arnaud2008}) as benchmark for the large observatory
scenario.

We choose the prominent Galactic black hole system and
microquasar GRS~1915+105 to see how the next missions could
investigate the GR effects on the degree and angle of polarization presented in
Sec.~\ref{sec:Polarization}. This source is very bright in the
thermal state and the inclination of 70$^\circ$ (Mirabel \& Rodriguez
\citeyear{Mirabel1994}) favours a high degree of polarization (see
Fig.~\ref{fig-pda1}).
We consider the spectrum in a low-luminosity thermal state (McClintock et al.
\citeyear{mccl06}), $L \lesssim 0.3\, L_{\rm edd}$, to minimize contamination
by a nonthermal component of emission.
The dependence of the
polarization with energy is that calculated in the Sec.~\ref{sec:Polarization}
with $\tau=1.0$ and $\theta_0=70^\circ$. The presented results are
computed for the hardening factor $f=1.7$, mass of the black hole
$\mbh=14\msun$ (Greiner, Cuby \& McCaughrean \citeyear{grei01})
and for the accretion rate $\dot{M}=1.4\times 10^{18}$ g/s
(McClintock et al. \citeyear{mccl06}).

We assume the performances of the current prototype
of the GPD simulated by means of a Monte Carlo software, which have
been essentially confirmed by recent measurements (Muleri et
al. \citeyear{Muleri2008}). The measured degree and angle of
polarization are derived from convolving the response of the GPD with
the expected degree of polarization and adding a Poisson background.

\begin{figure*}
\includegraphics{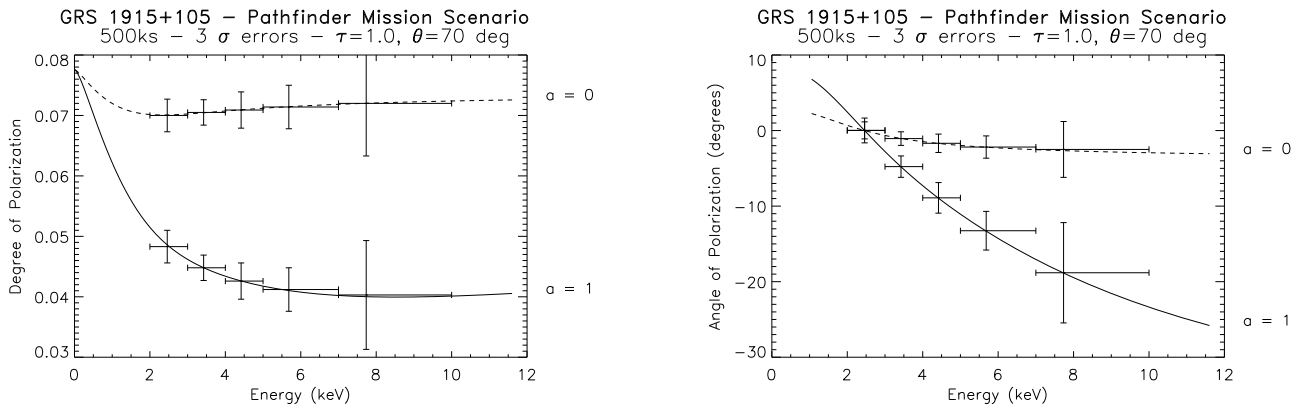}
\caption{Energy dependence of the degree (left) and angle (right) of
  polarization, measured with an observation of 500~ks by the
  pathfinder mission. The solid and dashed curves refer to the case of
  a Kerr and Schwarzschild black hole respectively, with $\tau=1.0$
  and $\theta_0=70^\circ$. Errors are quoted at the 3-$\sigma$ level. Note
  that we set the angle of polarization to zero at low energy;
  the actual offset of the observed angle depends on the
  inclination of the accretion disk, which is unknown at this stage.}
\label{fig:Pathfinder}
\end{figure*}

\begin{figure*}
\includegraphics{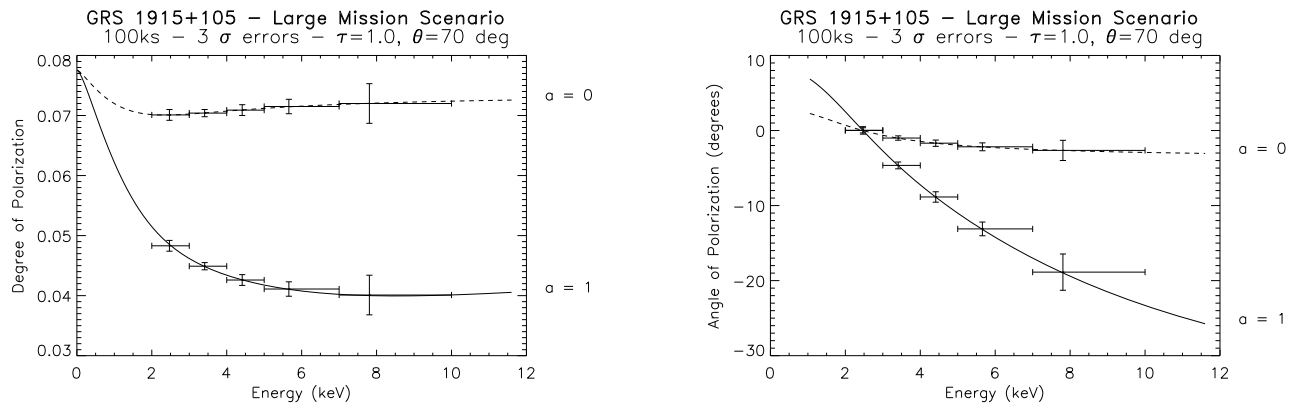}
\caption{The same as the Fig.~\ref{fig:Pathfinder} but for the large mission
scenario and an observation of 100~ks.}
\label{fig:LargeMission}
\end{figure*}

The results for the pathfinder scenario are shown in
Fig.~\ref{fig:Pathfinder} for an observation of 500~ks, while those for a
large mission are reported in
Fig.~\ref{fig:LargeMission} for an observation lasting 100~ks. The variation
with energy is easily observed for both degree and
angle of polarization. The Kerr and the Schwarzschild cases can be
clearly discriminated (the errors shown are at the 3-$\sigma$ level).

\section{Summary}

We have investigated the GR effects on the polarization properties of
the thermal radiation emitted by an accretion disc around a stellar
mass black hole. The polarization at infinity is changed from its local
value due to strong gravity effects and fast orbital motion of the disc
close to the central black hole. The difference is particularly
dramatic especially in the region below the critical point, where
the polarization angle swings over the entire range of 180 degrees.

The most interesting interval of energy is from 0.1 to 10 keV.
Below this energy band the emission comes
from far away from the disc centre, while above this range the flux
rapidly decreases. Note, however, that for different maximum temperatures
of the disc the value of the upper boundary of this energy
interval will change. All studied energy dependences will be shifted
to a higher energy for a hotter disc. The maximum temperature is a function of
the black hole mass, accretion rate and hardening factor as can be seen from
the eq.~(\ref{temp}).

The polarization degree is the highest for the lowest studied optical depth of
the disc's atmosphere and in most cases it grows with the observer's
inclination. The polarization
angle does not change much with the inclination of the observer apart
from the highest inclinations. Instead, it depends quite strongly on
the spin of the central black hole.

The effects of strong gravity on the polarization state of emergent
radiation will be accessible to the next missions with an X-ray
polarimeter based on the Gas Pixel Detector. We presented the case of the
pathfinder mission, feasible in a few years, and that of a large
mission. In both scenarios, the energy dependence of the degree and of
the angle of polarization will clearly discriminate between the Kerr and the
Schwarzschild black holes cases. Even in the pathfinder
scenario, the X-ray polarimetry will be a powerful probe of the metric
around black holes.

\section*{Acknowledgments}
The authors acknowledge E.~Costa, P.~Soffitta and
R.~Bellazzini for making available the information on sensitivity of the GPD
for the POLARIX and IXO missions.

This research is supported by the ESA Plan for Cooperating States
project No.~98040 in the Czech Republic. MD and VK
gratefully acknowledge support from the Czech Science Foundation grant
205/07/0052. RG is grateful for financial support to the Centre of
Theoretical Astrophysics (LC06014). FM and GM gratefully acknowledge the
hospitality
of the Astronomical Institute in Prague and, in particular, that of VK and MD.
FM also acknowledges financial support from Agenzia Spaziale Italiana (ASI)
under contract ASI~I/088/06/0.

{}

\label{lastpage}

\begin{thebibliography}{}

\bibitem[\protect\citeauthoryear{}{2008}]{Arnaud2008}Arnaud M.
et al.\ 2008, Experimental Astronomy, in press

\bibitem[\protect\citeauthoryear{}{2006a}]{Bellazzini2006a}Bellazzini R.
et al.\ 2006a,
NIMA, 566, 552

\bibitem[\protect\citeauthoryear{Bellazzini et
    al.}{2006b}]{Bellazzini2006b}Bellazzini R. et al.\ 2006b, Space
    Telescopes and Instrumentation II: Ultraviolet to Gamma
    Ray. Edited by Turner, Martin J. L.; Hasinger, G{\"u}nther.
    Proceedings of the SPIE, Volume 6266, p. 62663Z

\bibitem[\protect\citeauthoryear{}{2007}]{Bellazzini2007}Bellazzini R.
et al.\ 2007, NIMA, 579, 853


\bibitem[\protect\citeauthoryear{}{1998}]{belo98}Beloborodov A.~M.\ 1998,
ApJL, 496, L105


\bibitem[\protect\citeauthoryear{}{1960}]{chan60}Chandrasekhar S.\ 1960,
Radiative Transfer (New York: Dover)

\bibitem[\protect\citeauthoryear{Connors \& Stark}{1977}]{cs77}Connors P.~A.,
Stark R.~F.\ 1977, Nature, 269, 128

\bibitem[\protect\citeauthoryear{Connors et al.}{1980}]{con80}Connors P.~A.,
Stark R.~F., Piran~T.\ 1980, ApJ, 235, 224

\bibitem[\protect\citeauthoryear{Costa et al.}{2001}]{Costa2001}
Costa E., Soffitta P., Bellazzini R., Brez A., Lumb N., Spandre G.\ 2001,
Nature, 411, 662

\bibitem[\protect\citeauthoryear{Costa et al.}{2006}]{Costa2006} Costa E.
et al.\ 2006, Space Telescopes and Instrumentation II:
Ultraviolet to Gamma Ray. Edited by Turner, Martin J. L.; Hasinger,
G{\"u}nther. Proceedings of the SPIE, Volume 6266, p. 62660R

\bibitem[\protect\citeauthoryear{}{1975}]{cunn75}Cunningham C.~T.\ 1975,
ApJ, 202, 788


\bibitem[\protect\citeauthoryear{}{2004}]{dovc04a}Dov{\v c}iak, M.\
2004, PhD thesis, arXiv:astro-ph/0411605

\bibitem[\protect\citeauthoryear{}{2004a}]{dovc04b}Dov\v{c}iak M.,
Karas V., Matt G.\ 2004a, MNRAS, 355, 1005

\bibitem[\protect\citeauthoryear{}{2004b}]{dovc04c}Dov\v{c}iak M.,
Karas V., Yaqoob T.\ 2004b, ApJSS, 153, 205

\bibitem[\protect\citeauthoryear{}{2004c}]{dovc04d}Dov\v{c}iak M.,
Karas V., Matt G., Yaqoob T.\ 2004c, in Proc. of Workshops on Black
Holes and Neutron Stars, eds.\ S.~Hled\'{\i}k and Z.~Stuchl\'{\i}k
(Opava: Silesian University), pp. 33--73


\bibitem[\protect\citeauthoryear{Goosmann \& Gaskell}{2007}]{goos07}
Goosmann R.~W., Gaskell C.~M.\ 2007, A\&A, 465, 129

\bibitem[\protect\citeauthoryear{}{2001}]{grei01}
Greiner J., Cuby J.G., McCaughrean M.J.\ 2001, Nature, 414, 522

\bibitem[\protect\citeauthoryear{Laor et al.}{1990}]{lao90}Laor A.,
Netzer H., Piran T.\ 1990, MNRAS, 242, 560


\bibitem[\protect\citeauthoryear{Liu et al.}{2008}]{Liu}Liu J.,
McClintock J.~E., Narayan R., Davis S.~W., Orosz J.~A.\ 2008, ApJ, 679, L37

\bibitem[\protect\citeauthoryear{Matt}{1993}]{mat93}Matt G.\ 1993, MNRAS, 260,
663

\bibitem[\protect\citeauthoryear{Matt et al.}{1993}]{mfr93}Matt G.,
Fabian A.~C., Ross R.~R.\ 1993, MNRAS, 264, 839

\bibitem[\protect\citeauthoryear{}{2006}]{mccl06}McClintock J.~E., Shafee R.,
Narayan R., Remillard R.~A., Davis S.~W., Li L.-X.,\ 2006, ApJ, 652, 518

\bibitem[\protect\citeauthoryear{Miller}{2007}]{Miller07}Miller J.\ 2007
ARA\&A, 45, 441

\bibitem[\protect\citeauthoryear{}{1994}]{Mirabel1994}Mirabel I.~F.,
Rodriguez L.~F.\ 1994, Nature, 371, 46

\bibitem[\protect\citeauthoryear{}{2008}]{Muleri2008} Muleri F. et al.\ 2008,
NIMA, 584, 149


\bibitem[\protect\citeauthoryear{}{1973}]{novi73}Novikov I.~D., Thorne K.~S.\
1973, in Black Holes, eds.\ C.~DeWitt and B.~S.\ DeWitt (New York:
Gordon and Breach Publishers), p.~343

\bibitem[\protect\citeauthoryear{}{1974}]{page74}Page D.~N., Thorne K.~S.\
1974, ApJ, 191, 499

\bibitem[\protect\citeauthoryear{Pineault}{1977}]{pin77}Pineault~S.\ 1977,
MNRAS, 179, 691


\bibitem[\protect\citeauthoryear{}{1973}]{shak73}Shakura N. I., Syunyaev R. A.\
1973, A\&A, 24, 337

\bibitem[\protect\citeauthoryear{}{1995}]{shim95}Shimura T.,  Takahara F.\
1995, ApJ, 445, 780

\bibitem[\protect\citeauthoryear{}{2008}]{sila08}Silantev N.~A.,
Gnedin Y.~N.\ 2008, A\&A, 481, 217

\bibitem[\protect\citeauthoryear{Stark \& Connors}{1977}]{sc77}Stark
  R.~F., Connors, P.~A.\ 1977, Nature, 266, 429


\bibitem[\protect\citeauthoryear{}{1970}]{wal70} Walker M., Penrose~R.,\ 1970,
 Commun. Math. Phys., 18, 265

\bibitem[\protect\citeauthoryear{}{1978}]{Weisskopf1978}Weisskopf M.~C.,
Silver E.~H., Kestenbaum H.~L., Long K.~S., Novick R.\ 1978, ApJ,
220, L117


\end{thebibliography}
\end{document}